%% file: ms.tex
\documentclass{article}

\PassOptionsToPackage{numbers, compress, sort}{natbib}
\usepackage[final]{neurips_2023}
\usepackage[percent]{overpic}

\usepackage[utf8]{inputenc} % allow utf-8 input
\usepackage[T1]{fontenc}    % use 8-bit T1 fonts
\usepackage{hyperref}       % hyperlinks
\usepackage{hyperref}
\hypersetup{
    colorlinks=true,
    linkcolor=blue,
    filecolor=blue,      
    urlcolor=blue,
    citecolor=blue
}
\usepackage{url}            % simple URL typesetting
\usepackage{amsfonts}       % blackboard math symbols
\usepackage{nicefrac}       % compact symbols for 1/2, etc.
\usepackage{microtype}      % microtypography
\usepackage{xcolor}         % colors
\usepackage{enumitem}
\usepackage{authblk}

\usepackage{array}
\usepackage{amsmath}
\usepackage{booktabs}       % professional-quality tables
\usepackage{changepage}
\usepackage{multirow}
\usepackage{wrapfig}
\usepackage[small, compact]{titlesec}

\usepackage[font={small}]{caption}
\usepackage[export]{adjustbox}
\usepackage[skip=1ex]{caption}

\newcommand{\handlemath}[1]{\relax\ifmmode #1\else $#1$\fi}
\newcommand{\figref}[1]{{Figure~\ref{#1}}}
\newcommand{\secref}[1]{{\S\ref{#1}}}

\newcommand{\todo}[1]{}
\newcommand{\cjr}[1]{}
\newcommand{\ds}[1]{}
\newcommand{\dk}[1]{}
\newcommand{\ycx}[1]{}
\newcommand{\zh}[1]{}
\newcommand{\hf}[1]{}
\newcommand{\aditya}[1]{}
\newcommand{\swarat}[1]{}
\newcommand{\patrick}[1]{}
\newcommand{\nihal}[1]{}
\newcommand{\sriram}[1]{}
\newcommand{\daehyeok}[1]{}
\newcommand{\jane}[1]{}
\newcommand{\rohit}[1]{}
\newcommand{\zichao}[1]{}
\renewcommand{\todo}[1]{{\color{blue}\textbf{TODO:} #1}}

\renewcommand{\cjr}[1]{{\color{red}\textbf{[cjr]:} #1}}
\renewcommand{\ds}[1]{{\color{teal}\textbf{[ds]:} #1}}
\renewcommand{\dk}[1]{{\color{violet}\textbf{[donghyun]:} #1}}
\renewcommand{\ycx}[1]{{\color{brown}\textbf{[chenxi]:} #1}}
\renewcommand{\zh}[1]{{\color{green}\textbf{[zh]:} #1}}
\renewcommand{\aditya}[1]{{\color{pink}\textbf{[aditya]:} #1}}

\renewcommand{\patrick}[1]{{\color{cyan}\textbf{[patrick]:} #1}}
\renewcommand{\hf}[1]{{\color{purple}\textbf{[hf]:} #1}}
\renewcommand{\nihal}[1]{{\color{cyan}\textbf{[nihal]:} #1}}
\renewcommand{\sriram}[1]{{\color{orange}\textbf{[sriram]:} #1}}
\renewcommand{\daehyeok}[1]{{\color{blue}\textbf{[daehyeok]:} #1}}
\renewcommand{\jane}[1]{{\color{olive}\textbf{[jane]:} #1}}
\renewcommand{\rohit}[1]{{\color{yellow}\textbf{[rohit]:} #1}}
\renewcommand{\zichao}[1]{{\color{orange}\textbf{[zichao]:} #1}}

\newenvironment{denseitemize}{
\begin{itemize}[noitemsep, topsep=0pt, leftmargin=1.2em]
}{\end{itemize}}

\titlespacing*\section{0pt}{0pt plus 2pt minus 2pt}{0pt plus 2pt minus 2pt}
\titlespacing*\subsection{0pt}{0pt plus 2pt minus 2pt}{0pt plus 2pt minus 2pt}
\titlespacing*\subsubsection{0pt}{0pt plus 2pt minus 2pt}{0pt plus 2pt minus 2pt}

\makeatletter
\renewcommand\AB@affilsepx{, \protect\Affilfont}
\makeatother

\newif\iffinal
\finalfalse

\newcommand{\abbr}{FM4OS}

\title{On a Foundation Model for Operating Systems}

\author[1]{Divyanshu Saxena}
\author[1]{Nihal Sharma}
\author[1]{Donghyun Kim}
\author[1]{Rohit Dwivedula}
\author[1]{Jiayi Chen}
\author[1]{Chenxi Yang}
\author[1]{Sriram Ravula}
\author[1]{Zichao Hu}
\author[1]{Aditya Akella}
\author[2]{Sebastian Angel}
\author[1]{Joydeep Biswas}
\author[1]{Swarat Chaudhuri}
\author[1]{Isil Dillig}
\author[1]{Alex Dimakis}
\author[3]{P. Brighten Godfrey}
\author[1]{Daehyeok Kim}
\author[1]{Chris Rossbach}
\author[3]{Gang Wang}

\affil[1]{The University of Texas at Austin}
\affil[2]{University of Pennsylvania}
\affil[3]{University of Illinois at Urbana-Champaign}

\begin{document}

\maketitle
\input{texfiles/abstract}
\input{texfiles/introduction}
\input{texfiles/background}
\input{texfiles/fm4os}

\input{texfiles/usecases}
\input{texfiles/summary}

\bibliography{refs.bib}
\bibliographystyle{plainnat}

\input{texfiles/appendix}
%%%%%%%%%%%%%%%%%%%%%%%%%%%%%%%%%%%%%%%%%%%%%%%%%%%%%%%%%%%%

\end{document}

%% file: texfiles/abstract.tex
\begin{abstract}
This paper lays down the research agenda for a domain-specific foundation model for operating systems (OSes).
Our case for a foundation model revolves around the observations that several OS components {such as CPU, memory, and network subsystems} are interrelated and that OS traces offer the ideal dataset for a foundation model to grasp the intricacies of diverse OS components and their behavior in varying environments and workloads. 
We discuss a wide range of possibilities that then arise, from employing foundation models as policy agents to utilizing them as generators and predictors to assist traditional OS control algorithms.
% Throughout the paper, we identify key areas where integration of foundation models can lead to advancements in operating systems, while also discussing challenges and pitfalls one needs to be heedful of when designing these models.
Our hope is that this paper spurs further research into 
%By characterizing these opportunities, this paper hopes to inspire further exploration and innovation in developing an 
OS foundation models and creating the next generation of operating systems for the evolving computing landscape.
% \aditya{we should use a singular "foundation model" above and below? Because we are proposing one model.}
\end{abstract}

%% file: texfiles/introduction.tex
\section{Introduction}

%%\begin{figure}[hbp]
\begin{wrapfigure}{r}{.6\linewidth} 
    \centering
    \vspace{-6mm}
    \includegraphics[width=\linewidth]{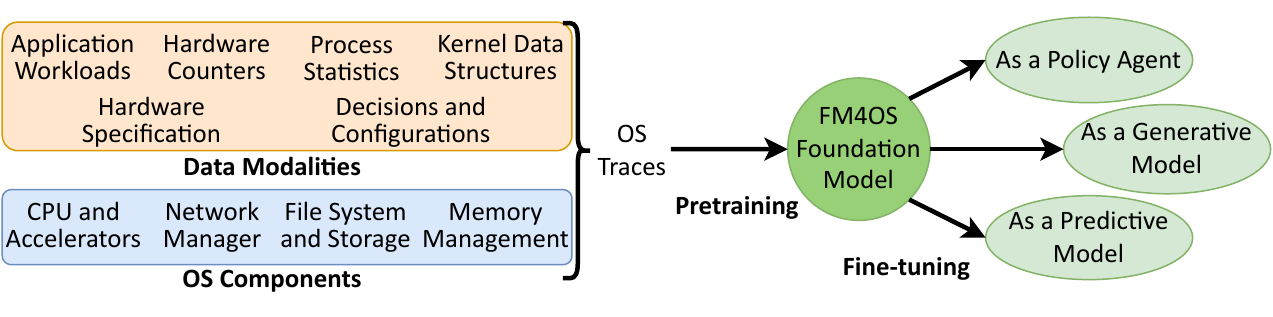}
    \caption{\small \abbr{}: a foundation model for operating systems.}
    %%\vspace{-3mm}
    \label{fig:workflow}
    \vspace{-2mm}
%%\end{figure}
\end{wrapfigure}

The Operating System (OS) is the central pillar of modern computing systems, overseeing hardware and software resources and enabling applications ranging from assistive robotics to cloud services.
OSes serve vital tasks such as scheduling processes; managing CPU, network, and memory resources, and interfacing with devices.
To make good decisions, OS policies must account for complex system dynamics such as hardware variances and environment responses, which is challenging for two reasons.
First, OSes can be deployed atop a variety of hardware, and amidst diverse workloads and environments.
Second, the OS does not have full visibility of the environment (e.g., network performance) or the workload (e.g., application request patterns), making the state space \textit{uncertain}. 
% \aditya{is the first talking about environment uncertainty and the second about state space uncertainty? can we make that clearer? or perhaps I'm misreading this and the two points are referring to something else.}

Conventional OS policies, reliant on manual algorithms or heuristics, lack adaptability across hardware, environments, and workloads, and often require manual tuning. 
% This is evident in edge servers struggling to run real-time applications close to the users, and cloud servers where operators find it challenging to ensure high performance and high utilization.
Recent proposals for using machine learning (ML) models in OS components such as the network manager~\cite{orca-sigcomm-20,rl-icml-19}, memory manager~\cite{cachesifter-fast-22,parrot-jmlr-20,lrb-nsdi-20,glcache-fast-23} and CPU scheduler~\cite{mllb-sigops-20}, while being good starting points in bringing data-driven decisions, are still far from ideal as they only optimize for individual components. Furthermore,  they neither integrate well together nor generalize well for diverse environments.

Inspired by the recent successes of large unsupervised ``foundation'' models in NLP and vision tasks, we argue that it is time for the OS to eschew such task-specific solutions in favor of foundation models.
\iffinal
\else
We envision the use of foundation models in OSes to present a significant step in the evolution of the field of OSes -- architecturally, an OS foundation model can replace a plethora of handcrafted heuristics in OSes with data-driven decisions that automatically adapt, and they can fundamentally transform the implementation, management, and operations of OSes.

\fi
% \sriram{We may want to re-consider the phrase ``end-to-end'' - in ML research, this implies a task-specific model trained in a supervised fashion}.
% These foundation models are first pretrained over a large dataset in an unsupervised fashion and then fine-tuned for various downstream tasks.
% \todo{say what OS traces are and how they capture all that is needed.}
Our insight is that OS traces consisting of hardware metrics, system event logs, and application arrivals and requests, can capture all the information on the workings of various OS components and the impact of their decisions on each other.
Further, OS traces collected on diverse hardware and application workloads can also capture the intricate relationship between OS decisions, hardware features, and application workloads.
We argue that a foundation model trained on such traces, \abbr{}, is plausible and can be used for several downstream tasks (as shown in \figref{fig:workflow}).
\iffinal
\else
In this paper, we discuss how \abbr{} can be trained and fine-tuned and the research challenges that arise.
\fi

% \ds{end -- edited.} \aditya{since we don't actually build the FM, I toned this down a bit}
% the various OS components have overlapping state spaces, meaning that we can combine the otherwise separately processed state spaces for task-specific models into a single input space for a foundation model.
% We argue that the various interrelated OS components and the dynamics of the OS environment (i.e., hardware resources and application workloads) are very complex, making the OS a rich arena for a domain-specific foundation model.

% In this paper, we first provide background on OSes and foundation models in \secref{sec:background}, and motivate the development of a new domain-specific foundation model in \secref{sec:fm4os}. We then provide a systematic analysis of various ways in which such a foundation model can be employed in an OS and the fundamental research challenges that arise in both computer systems and machine learning.

%% file: texfiles/background.tex
\section{Background}\label{sec:background}

In this section, we first provide background on OS decision-making: what makes it difficult and why adaptive decisions are needed, and then we give a brief background on foundation models. %before motivating the use of a foundation model for OS via a concrete example.

\noindent\textbf{Desiderata for operating systems.}
Operating systems oversee hardware and software resources, including CPU, memory, storage, and network (Table~\ref{tab:os_components} in appendix).
In general, OS tasks can be considered sequential decision-making processes where past actions and states of the OS instruct the action at any time. However, these can be very complex because:

\iffinal
\vspace{-2mm}
\fi

\begin{denseitemize}
    \item OSes can be deployed on diverse hardware with differing performance profiles. Further, they can run different workloads (e.g., microservice~\cite{alibaba-uservice-21} vs. ML workloads~\cite{mlaas-nsdi-22}) with varying objectives (e.g., prioritize power efficiency for robots vs. optimize performance for cloud servers).
    \item Access to fine-grained metrics (like the ones shown in Table~\ref{tab:os_components}: System State column) from hardware devices or the OS kernel, may be limited.
    \item \textit{System dynamics}, i.e., the interplay of policies between OS components, also plays a role in decision-making because the actions of one component can impact the future states of other components. Capturing these intricate dynamics is difficult due to the myriad OS policy combinations.
\end{denseitemize}

\iffinal
\vspace{-2mm}
\fi

Thus, in the OS setting, there is an \textit{inherent uncertainty  and partial observability in the state}.

\textit{Existing methods:}
Prior research has proposed learned and data-driven approaches to address these challenges. Some have employed DNNs to learn policies for specific OS components~\cite{mllb-sigops-20, linnOS-osdi-20, llama-asplos-20, sinan-asplos-21, kml-hotstorage-21, learned-os-19}  
while others have tackled state uncertainty by modeling OS tasks as MDPs~\cite{orca-sigcomm-20, park-neurips-19, lecar-hotstorage-18, glcache-fast-23, firm-osdi-20}. % and using RL methods.
% \aditya{there are non RL ones too.. do they belong in a different bin?}
Additionally, statistical and deep learning methods have been explored to generate realistic workloads~\cite{sdgen-fast-15, netshare-sigcomm-22, doppleganger-imc-20, cloudrnn-sosp-21, surakav} that can help inform conventional policies.
However, these approaches remain \textit{point solutions} that model individual OS components, leading to a diverse bag of policies, operating independently of others. Consequently, they fall short in optimizing end-to-end OS performance and decision-making.
\textit{Ideally, if we could learn how an OS task is impacted by other OS components, application workloads, and hardware specifications, we can devise methods to optimize OS decisions for desired objectives.}
These existing approaches also struggle with generalization beyond their training distribution, as shown in prior research~\cite{distribution-neurips-20,whirl-sigcomm-21}.
\textit{Therefore, we need techniques that generalize well to unseen inputs.}
% \textit{Recent works have shown that such supervision biases can be tamed by pre-training large unsupervised `foundation' models}~\cite{masked-arxiv-21}, as described below. \rohit{Is ``supervision bias" a well known term within the ML/LLM world?}~\ds{I picked it from the climax paper -- Nihal/Sriram: is this okay?}

% ~\cjr{We don't cite our own LAKE paper--is the because it took no position on what ML techniques should be used?}~\ds{yes, LAKE does not give a new ML technique but shows how to incorporate existing techniques in the kernel.}\aditya{I think we can cite LAKE with the last sentence as we do show limitations of existing learned approaches. I added a ref.}

\noindent\textbf{Foundation models.} This is a catch-all term for ML models trained on a large and diverse dataset to understand the general structure of the data and then fine-tuned (with much less data) for specific tasks.
While they have been touted as useful in myriad settings \cite{foundation-opportunities-22, foundation-decisions-23}, these models have shown tremendous empirical success in sequence modeling problems in natural language~\cite{chowdhery2022palm, driess2023palme, gpt-language-20}, finance~\cite{bloomberggpt-foundation-23}, computer vision~\cite{stablediffusion-cvpr-22}, biomedical imaging~\cite{generalist-biomedical-23} and climate modeling~\cite{climax-icml-23} to name a few.
At the core of these successes is efficient use of the transformer architecture~\cite{transformer-nips-17}, which learns long-term (spatial and/or) temporal correlations between input sequences, and the principles of transfer learning~\cite{transferlearning-98}, that enable learning for different tasks, domains, and modalities.

Several OS tasks also fall into this broad category of sequential modeling with the important caveat that the cadence with which decisions are made and the amount (time) and explicit form (states) of past observations vary widely between tasks (see Table \ref{tab:os_components} in Appendix~\ref{apdx:mdps}).
% Intuitively, having one model that captures the nuanced relationships in the system dynamics could be used to address this diverse set of tasks.
% Simply using existing foundation models trained on language or image data is not enough as they are unsuitable for understanding relationships between floating point numbers representing various OS-specific information, especially when the data is temporal (see~\cite{pitfalls-llm-23} for some evidence).
\iffinal
\else
The aforementioned successes of foundation models for tasks that share a common `foundation' lead us to ask: Can a foundation model built for the OS encode the correlations in the shared state space and impacts of hardware configurations and workloads to potentially improve downstream decision-making?
\fi
% \dk{This para only deals with how different OS components can be learned together through foundation models. Would it be Ok not to mention about different OS environments/hardware specs here?}\ds{edited -- ptal.}

%% file: texfiles/fm4os.tex
\section{\abbr{}: A Foundation Model for the OS}\label{sec:fm4os}

We propose the development of \abbr{}, a foundation model that understands the ``natural behavior'' of the OS and can be fine-tuned for several classes of downstream tasks - all of which either replace or aid the existing policies in the OS.
% We discuss why such a foundation model makes sense \aditya{"makes sense" is vague. Also, I am not sure how the SCH example below is useful. Isn't it better to (1) say something about detailed logs in systems today first, all the sources that can be combined together and then broefly say that those encode many things, not just for SCH but also others. (2) Use the specific example of SCH to say what all about it are encoded in the trace. (3) then say briefly about a model trained on this trace can be used for SCH-related things as you do toward the end of the next para} and what it would look like. \aditya{in this second part you should talk about why this is a FOUNDATION model. Say here that (3) other tasks such as XX have shared state space components with SCH and others, and that they are not entirely independent with each other. Move the specfics about dependence between SCH and others over here or talk about it here again.}
We begin by describing the data sources available that can be used to train such a model. %, and then discuss why it is a `foundation model' by giving a concrete example for the OS scheduling task.

\textbf{Data Sources.} Today's OSes, along with associated monitoring and data collection infrastructures, provide data in several forms (as shown in \figref{fig:workflow}), including
\iffinal
\textit{logs from OS components, hardware metrics, and application workloads}. We elaborate on these sources in Appendix~\ref{apdx:sources}.
\else
\begin{denseitemize}
    \item \textit{Action logs from OS components:}  Kernel logs such as {\tt dmesg} in Linux/MacOS and event logs in Windows, 
    %when configured during boot, 
    capture kernel debugging data, hardware events (e.g., network link status), and system events like interrupts, process restarts. These logs capture the actions taken by the OS components and the system state used by OS tasks (System state and Actions columns in Table~\ref{tab:os_components}).
    % ~\cjr{Is it worth noting that dmesg is present in Linux and MacOS, but other OSes by necessity have similar tools (e.g. Windows has the Windows System Log.}
    % We can also get similar statistics on system calls and kernel data structures using eBPF~\cite{ebpf}.
    \item \textit{Resource metrics and hardware counters:} Hardware drivers record several quantities relating to the resource's state at a pre-configured frequency. These include CPU, memory, and disk bandwidth utilization, NIC queue length, and hardware counters. 
    %Frequently collected metrics from the hardware drivers about the state of the resource (e.g. CPU utilization, memory usage, disk bandwidth usage, NIC queue lengths, CPU hardware counters, etc.)
    \item \textit{Application workloads:} Workload traces from productions~\cite{alibaba-uservice-21, serverless-azure-20, borg-eurosys-15}, public infrastructures~\cite{taccstats} and synthetically generated ones offer detailed application-level information, such as application type, request arrival rates, statistics of resource usage during execution.
    % \item \textit{OS environment metadata:} This includes information about the hardware specifications \aditya{is it static info, or counters and other dynamic info like that?}\ds{mostly static -- but we need to collect traces on several environments is the point} of the system resources (e.g. CPU, Cache, RAM, NIC, file system, etc.) and the deployed environment (i.e. cloud server, robot, or edge server). This data directly reflects the Environment State column for the different OS tasks in Table~\ref{tab:os_components}.
\end{denseitemize}
\fi

We will use the term ``OS trace'' to refer to the union of the data corresponding to a single machine drawn from the sources above, represented as a single (time-annotated) sequence. Such traces can be collected from systems with varying hardware specs (CPU, Cache, RAM, NIC, file system, etc.) and under various deployments (cloud, robots, and edge). 
%\aditya{drop the next sentences. Doesn't add much? Or roll it into previous bullets.} Information in the traces about the hardware spec and deployment scenario is the OS environment metadata, that directly reflects the Environment Information column in Table~\ref{tab:os_components}.
% \aditya{drop the rest of this para. Repeats from the above}
% An OS trace encodes all the relevant information needed for task-specific decision-making (as shown in Table~\ref{tab:os_components}: system state, environment state, and action columns).
% System state comes from various OS component logs and environment metadata, policy actions are recorded in the logs, and environment state is reflected in the application workloads.
% Further, by consolidating time-series information, the OS traces also encode causal correlations of workloads and past OS decisions on the resulting states.
Below, we describe two OS tasks that operate on different parts of the system, both of which can be trained from the OS traces. 

% Below, we describe how these OS traces can be used to train a model for one such OS task, \textbf{SCH}, and then show that these traces are rich enough to be used to train a foundation model for various downstream tasks, not limited to the \textbf{SCH} OS component.\nihal{Dangerous; this reads more like transfer learning than foundation models. Replace full paragraph with:}

\noindent\textbf{An example use case. } 
Consider the OS scheduling task \textbf{SCH} and cache replacement task \textbf{CACHE} (descriptions as given in Table~\ref{tab:os_components}).
% For the scheduling task: the goal is, given a queue of processes and $n$ CPU cores, assign a process to each core for the next time quanta (say, 100ms). For each time quanta, the scheduler pauses the ongoing process, adds it back to the queue (if the process hasn't already finished), and repeats the process.
As shown in the table, optimal decision-making for the \textbf{SCH} task requires process states, process completion times, hardware state, and process arrival workloads while the \textbf{CACHE} task requires cache size, state, and cache access workloads. 
All of these are captured in the OS traces. For \textbf{SCH}, the process states and completion times are captured in the logs, process arrival workloads and hardware aspects are captured by the application workloads and environment metadata, respectively.
Similarly, for \textbf{CACHE}, the cache state is captured by the resource metrics, cache size by environment metadata, and cache access patterns by application workloads.

The OS traces also capture the relationships between the two tasks.
For example, the process completion times would depend on the hardware specs of resources other than CPUs, such as the cache. This is because processes may access resources other than CPUs during their execution.
For the same reason, OS decisions relating to \textbf{CACHE} would also impact the process completion times.
Since our OS traces record features that cover the input space of both tasks \textbf{SCH} and \textbf{CACHE}, they can be used to train one model that can orchestrate both.
This model can then be used for several downstream tasks, including: (i) directly making good-quality decisions for the \textbf{SCH} and \textbf{CACHE} tasks, (ii) predicting the completion time of a newly arrived process, or (iii) generating traces for \textbf{CACHE} tasks that can be used to improve conventional data-driven or ML-based algorithms.

% the CPU (max frequency, cache size, bus speed, etc.) \textit{as well as} other resources, such as cache, RAM, network cards, and storage disks - this is because processes may access resources other than CPUs in the course of their execution. All of these are captured in the environment metadata.
% For the same reason, OS decisions relating to components other than CPUs (e.g. \textbf{CACHE} decisions for cache replacement, \textbf{CC} decisions for congestion control) also impact the process completion times - these are also captured in the logs. \aditya{flow and writing can be improved/tightened in this para}

% For this task, we can get information about scheduling events and the state of scheduling data structures, details of cache replacement, and congestion control tasks via {\tt dmesg} logs (used for debugging the kernel).
\iffinal
When trained on diverse OS traces, the model learns how scheduler and cache behaviors relate to hardware and workloads, enabling generalization to predict program performance on new CPU specifications and cache sizes.
\else
Additionally, when trained on OS traces spanning various hardware and workload combinations, the model would grasp how scheduler and cache replacement behaviors relate to hardware specs and application workloads. This enables generalization to unseen scenarios, including tasks like predicting program performance on new CPU specifications and cache sizes.
\fi
% would understand the behavior of the CPU scheduler, with respect to the hardware, application workloads, and the choice of algorithms used in other OS components.

% \nihal{@Divyanshu, this paragraph needs reformatting: Sch task is <blah>, uses <blah> to make optimal decisions. Cache task is <blah>. While it shares <blah> with sch to make decisions, it also requires <blah> to take actions. Since our OS trace data records features that cover the input space of both these tasks, it can be used to train one model that can do both. 
% While these are only two example tasks, almost all tasks in the OS present with such shared inputs. This leads us to believe that there is scope for the development of a foundational model in the space of OSes.}
% \dk{I somewhat feel like the above example does not show the need for collecting OS traces from different hardware combinations.}
% \ds{Nihal, Donghyun: Edited -- ptal.}

\noindent\textbf{Foundation Model for the OS.}
The OS traces used above not only encode information for \textbf{SCH} and \textbf{CACHE} tasks {\em but also that corresponding to the decision-making in several other OS components}, e.g., I/O prefetching, packet scheduling, congestion control policies, etc.
\iffinal
Referring to Table~\ref{tab:os_components}, we make two observations to support this. Firstly, several of the OS tasks have shared state space components. For example, both \textbf{PREFETCH} and \textbf{CACHE} tasks need the cache state, both \textbf{PREFETCH} and \textbf{PAGE} tasks require process instructions, etc.
\else
Referring back to Table~\ref{tab:os_components}, we make two observations to support this argument. Firstly, several of the OS tasks have shared state space components. For example, both \textbf{PREFETCH} and \textbf{CACHE} tasks need the cache state, both \textbf{PREFETCH} and \textbf{PAGE} tasks require process instructions, etc.
\fi
Secondly, these tasks are not entirely independent, as shown in the above \textbf{SCH} task example, where the process completion times (needed for \textbf{SCH}) depend on the policies adopted in the \textbf{CACHE} task.
This inter-dependence of one component on others is a widely seen and natural phenomenon in the operating system.
% \aditya{drop this: Therefore, OS traces collected from different OS components are very rich because even the events corresponding to one task (e.g. \textbf{CACHE}) capture information about other tasks. }
\iffinal
Using
\else
These two observations lead us to posit that, using
\fi
OS traces collected across many machines, one can therefore build a foundation model -- \abbr{}, that knows the `natural behavior' of the OS.
\iffinal
A prospective pretaining regime for \abbr{} is discussed in Appendix~\ref{apdx:pretrain}. 
\fi
% \dk{To my understanding, \abbr{} solves two difficult problems together: 1) considering all components together, 2) learning OS behaviors varying hardware/environments.
% For me, it is a bit unclear how \abbr{} can solve the two problems at once. 
% Especially, "Desiderata for Operationg Systems" para in background section covers the importance of understanding various system environments.
% But, section 3 rather focuses more on how \abbr{} can understand OS components together in one model.
% }\ds{edited background and example usecase to hint at the varying hardware -- ptal.}

\iffinal
\else
We envision \abbr{} to be {\bf pre-trained} using self-supervised methods on a large corpus of OS traces. This pre-trained model would capture temporal relationships in the sequence of inputs it accepts and build an understanding of the system dynamics of the OS.
Existing literature (particularly in natural language) has proposed several pre-training tasks that can be used to develop this basic understanding. Notable among these are the Next Token Prediction~\cite{gpt-18}, Masked Language Modeling, Next Sentence Prediction~\cite{bert-18}; each with their own pros and cons.
While it seems intuitive to employ an auto-regressive model, pre-trained with next token prediction to build \abbr{}, the optimal pre-training task is an open and interesting question in itself.
\fi

%% file: texfiles/usecases.tex
\section{Downstream Tasks for \abbr}
 % \begin{figure}
 %     \centering
 %     \includegraphics[width=0.7\textwidth]{figures/usecases-chart.pdf}
 %     \caption{Usecases of Foundation Models in Operating Systems}
 %     \label{fig:usecases}
 % \end{figure}

% \aditya{would be good to have a figure showing/exemplifying these tasks}

We are now ready to discuss the fine-tuning of \abbr{}. We present key downstream tasks and categorize them into three broad use cases: as a policy agent, a generative model, and a predictive model.
We discuss these individually below and highlight challenges unique to the OS setting that require novel research on training and using
\iffinal
foundation models, in Appendix~\ref{apdx:challenges}. 
\else
foundation models.
\fi
% \todo{for each downstream task, write how the task can be achieved - via fine-tuning or directly, etc.}

\subsection{\abbr{} as a Policy Agent}
% \todo{(Rohit: am working on this)}
As discussed in \secref{sec:background}, several OS tasks can be modeled as a sequential decision-making process where the state of the OS evolves according to the actions a policy makes. Prior works~\cite{firm-osdi-20,orca-sigcomm-20,mllb-sigops-20,linnOS-osdi-20} have used handcrafted features based on heuristics in order to model complex system dynamics. 
% and the requirement for different features is also dependent on the environment
% under which the system is running. For example, the energy consumption of a particular
% component may not be a feature in the cloud OS, but it is required to make good decisions
% in a robot OS. \aditya{the fact that you are considering OSes for different settings, e.g., robots, cloud, edge, did not come through clearly in earlier sections.}

% Given the recent advances in foundation models and their capabilities in transforming
% a large feature space into a smaller latent space ~\cite{} \todo{add citations for
% stable-diffusion type models, GPT-2 type too?}, we believe it is worth exploring the
% use-case for a foundation model acting as the agent `in-the-loop'. Such in-the-loop
% foundation models can be classified into two broad categories, as described below.

The key challenge for any solution addressing multiple tasks in the OS is the diversity in state and action spaces of tasks and the different lengths of temporal history deemed relevant for each task (see Table \ref{tab:os_components}).
Foundation models have been shown to solve precisely this issue of varying lengths of temporal history due to their ability to summarize inputs of arbitrary lengths in a common representation space.
Further, they have also shown evidence of being capable of handling multi-modal input data~\cite{multimodal-foundation-23}, which suits the various forms of information captured in OS traces (see \S\ref{sec:fm4os}). By engineering the size of these representations for pre-training and specifying the objective during fine-tuning, we expect that \abbr{} can be used to suggest optimal actions.
% (see \todo{cite gpt2, T5, other works that show generalizability} for evidence of this behavior in other domains).

%Dealing with the complicated system dynamics, huge state space, and action space for OS tasks requires models with good representative power \ds{can we get some citation here?}. The power of foundation models is that they can be adapted to a wide range of tasks when trained once on enough data. With the development of \abbr{}, we would like to understand if we can train a foundation model to understand system dynamics and use it as the `in-the-loop' agent making decisions.
% Foundation models, however, are not omnipotent; foundation models have a limited set of tasks \todo{Add citation} and should be trained on corresponding training data \todo{Add citation}. 
% \nihal{Claims of huge state and actions spaces need backing either in numbers or in citations with similar  claims.}

% \aditya{"task" is used in a confusing manner. Both to describe downstream tasks, but also inputs to scheduling algorithms.}\nihal{please reread to see if edits solve this}

\noindent\textbf{Making low-level decisions:} By pre-training \abbr{} over OS traces, we expect it to understand the semantic space for OS decision-making.
\iffinal
\else
This encompasses trade-offs between performance and resource usage, component relationships, and system properties. These semantics remain consistent across different OSes regardless of changes in workload, hardware, or environments.
\fi
Then, we can use \abbr{} to take low-level actions for OS tasks, such as setting the congestion window for the \textbf{CC} task and choosing processes for the \textbf{SCH} task.
Fine-tuning to make these decisions requires historical traces labeled with optimal actions.

\noindent\textbf{Policy selection:} Current inference times for transformer-based models do not match the pace at which some OS tasks require actions (every few $ns$). Accelerating inference~\cite{flashattention-neurips-22, deebert-acl-20}, especially for operation in the OS~\cite{lake-asplos-23} is an ongoing research area.
In the meantime, \abbr{} can address the relatively simpler task of selecting from existing policies (over longer time frames) instead of specifying actions explicitly.
For each task, there exist policies optimized for specific environments and workloads. For instance, for the \textbf{CACHE} task, Least Recently Used (LRU) policy is favored when access patterns follow locality trends, while Least Frequently Used (LFU) policy is more suitable for random accesses with consistent popular requests~\cite{vietri2018driving, rodriguez2021learning}. 

\iffalse
\noindent\textbf{Generating system code:} In a more abstract sense, any policy is simply a block of code operating in the OS kernel. Using this definition of a policy also provides the additional benefit of being able to explain and interpret the actions suggested by the model.
Recent work in ~\cite{liang2023code, codebotler2023} has shown evidence of a policy that calls various APIs to generate OS code. 
Further, language models have also been capable of generating code \cite{chen2021evaluating}. We can define a class of language model programs (LMP) over different OS applications by providing few-shot examples. For example, we could provide some textbook examples of generating device drivers and then ask the foundation model to help us generate drivers for new devices.
% \aditya{for what code specifically? heuristic code, or any OS code?} 
\nihal{This paragraph can go. Seems too complex. Going from language->code is okay. But our model is trained on traces. How are we prompting it to output code? Current explanation is really hand-wavy.} \aditya{I agree. This does not seem related to the main story.}
\fi

\iffinal
\else
\noindent\textbf{Challenges:} End-to-end application performance depends on collective decisions made by OS components.
Using foundation models as policy agents brings two unique challenges:  \textit{composability} of actions from various policy agents and end-to-end \textit{explanability} of their decisions.
The former arises because decisions of one policy can affect the future states of other agents (as with the \textbf{CACHE} and \textbf{SCH} example discussed in \secref{sec:background}).
Independently fine-tuned components in the OS may result in suboptimal OS-wide decisions, that may affect both individual application and system-wide guarantees (e.g. fairness and starvation-freedom). 
% On the other hand, system-wide guarantees on decision quality and explanations for individual decisions necessitate a holistic understanding of OS behavior.
One possible approach here is to jointly fine-tune components (to ensure concerted decisions) as well as to develop techniques that provide component-wise guarantees (on performance, e.g., bounds on tail request completion times, or correctness, e.g., safety and liveness properties~\cite{whirl-sigcomm-21}), and \textit{formally guaranteed composability} of these actions to provide global invariants for the entire OS.

Regarding the latter, ideally, we desire human users to understand the OS at some level to audit or debug it. However, learned decisions from a black-box model may easily obscure the understanding of overall behavior.
We envision the use of approaches that describe what each learned policy did in a given execution (similar to LIME~\cite{lime-2016}), what could have happened had a learned policy made a different decision, and also produce human-comprehensible `summaries' in the form of rules~\cite{dillig-counterfactual-expl-icse-22,dillig-explaining-mispredictions-2021}, or programs~\cite{interpretable-icml-18-swarat} of what the module will do ahead of time.
\fi

\subsection{\abbr{} as a Generative Model}

Content generated by ML models offers new opportunities for OSes, similar to benefits observed from using generative models in other domains~\cite{cloudrnn-sosp-21,stablediffusion-cvpr-22}. Synthetically generated data can help add diversity to existing training data used by data-driven solutions, help with the availability and sharing of proprietary or confidential data, and testing models under settings that occur infrequently in practice. 

%overcoming key practical concerns such as intellectual property concerns (more below) [daehyeok]: what are the concerns?. [aditya]: PTAL? 
%It can also help provide a new avenue for testing OS functionalities against a much wider variety of realistic settings than possible today. %We elaborate on these below.
% \aditya{rewrote this. Is this what you meant to say for the second part? }

%; crucial, such synthetic data can preserve privacy-sensitive inputs yet be of great utility. In addition, it can also assist in analyzing complex data and finding hidden patterns in new ways that raw data may not reveal. \aditya{I don't understand this last part. Drop it?}
% \check{do we need foundation models for this task or are transformers (or task-specific generative models) enough?}

\noindent\textbf{Generating traces:} The lack of (diverse) training data is a major hurdle in most data-driven and learned approaches for OS tasks.
Even if such data were available, storing and maintaining such a large corpus of data collected under different hardware configurations and workloads be challenging.
For example, for the \textbf{CACHE} task, traces are needed for different memory specifications and for different types of workloads (small objects, large objects, mixed sizes, etc.).
\iffinal
\else
For tasks that require network information, large volumes of data remain in the hands of corporations that restrict access due to intellectual property and confidentiality reasons. 
\fi
% For instance, public traces for scheduling tasks (e.g., google, alibaba traces) include desensitized information such as user and process information, which can be helpful to find workload patterns.
By training \abbr{} using auto-regressive tasks like Next Token Prediction, we could train the model to learn to generate OS traces that can be used in a variety of ways.

Fine-tuning it with specially designed prompts could lead to traces that adhere to specific constraints (e.g., setting hardware configurations, limiting network bandwidth, etc). These can be used to supplement the training data collected on specific configurations.
Further, the foundation model can also be fine-tuned to obfuscate confidential information from the traces while keeping the important relationships of the traces intact (prior works~\cite{netshare-sigcomm-22, doppleganger-imc-20} show feasibility of such obfuscation in network traces). 
Another opportunity that we identify here is that \abbr{} can also be used to generate \textit{pathological corner cases}. Specifically, we posit that by appropriately querying the foundation model, we can use it to generate {\em pathological corner cases} that would have been otherwise difficult to get.

\iffinal
\else
\noindent\textbf{Challenges:} As with any generative model, quantifying the quality of synthetic samples is a key challenge.
At the very least, these traces should
% closely match real trace distributions (e.g., request arrival and size patterns for a \textbf{CACHE} task trace must be similar to real-world traces) while maintaining
maintain certain relationships between variables (e.g., total network transmissions should be less than network bandwidth).
They must also capture desired properties that are difficult to obtain otherwise, such as `realism', i.e., a specific sequence of requests in a generated trace can actually arise in practice --- this is an avenue for future research.
Further, the generated traces should also be diverse to be useful. For example, for a cache replacement algorithm, we would want traces with diverse and realistic combinations of small and large object arrivals to effectively stress-test the algorithm~\cite{darwin-sigcomm-23}. 
% \dk{A sentence explaining the privacy challenge is gone. If it was intentional, I think we should remove the following sentences also.}\ds{I merged that in the following one - the following sentences capture the privacy challenge, no?}
Another challenge is with leakage and memorization of sensitive data. As shown in previous works~\cite{diffusion-sec-23,codex-sec-23}, carefully designed prompts can extract memorized training data with sensitive information. Thus, integrating techniques such as filtering the memorized data~\cite{llm-sec-21} and ideas from say, differential privacy~\cite{dp-tcc-06}, into \abbr{} are necessary.
\fi

%\aditya{can we say a bit more? Is this related to DP?} \dk{As revealed in previous works~\cite{llm-sec-21,diffusion-sec-23,codex-sec-23}, carefully designed input prompts can extract memorized training data with sensitive information. To minimize privacy leakage, use of differential privacy~\cite{dp-tcc-06} should be considered. Or we can try filtering out memorized output as ~\cite{llm-sec-21} suggested as a future work. When mitigating privacy leakage issues, we should also take care not to hurt generation accuracy.}
% This requires us to `query' the generative model appropriately. We vision a DSL to query such a generative model, where the DSL can capture properties about the traces or testcases that we want to generate and then the model can use the DSL query to generate the required traces.
% Coming up with such a general-purpose DSL is a key challenge.

\subsection{\abbr{} as a Predictive Model}
Foundation models have been shown to exhibit good performance on downstream prediction tasks~\cite{gpt-language-20,climax-icml-23}. 
In the OS setting, we can use \abbr{} as an encoder of the state, and then use linear probing to predict various things about the system's response, future utilization. This can lead to efficient placement, scheduling, performance, and anomaly detection. %It can also be used to build more robust OS heuristics and algorithms~\cite{cloudrnn-sosp-21, madu-socc-22}.
% \aditya{is it a requirement, or can it help build better more robust systems?}

\noindent\textbf{System response prediction:}
Understanding how the environment of the OS evolves with application-level decisions made by the OS are crucial to improve decision quality.
For example, predicting the time to completion of a process would allow the kernel to reorder its CPU work queue based on completion times leading to an optimal schedule for minimum waiting time of jobs.
Since \abbr{} is pre-trained to understand precisely the needed semantic relationships between OS subsystems, it can be used to closely predict system responses.
% We note that this is essentially a summarized version of its operation as a generative model. 

%For example, for a CPU scheduling component, the completion time for a process depends on how the instructions are processed on the hardware and how the various memory and storage components respond. \aditya{this isn't clear} Accurately predicting the completion times for the next $N$ processes would allow a traditional kernel to reorder them in the shortest completion time; if the prediction is accurate, the schedule will be optimal!

%The environment of the OS consists of all the external artifacts that affect its performance,  for example, the response of hardware resources, like PCIe buses, network cards, memory cards, accelerators, etc.
%How this environment responds to an application and the decisions an OS takes for the application, and how the application performance is affected as a result, are crucial for making good quality decisions in the OS (as shown in Table~\ref{tab:os_components}). \abbr{} can play an important role in capturing these:
%\abbr{} can be fine-tuned to predict the performance of the process given a specific configuration of the system resources (i.e., the current environment); 
%this performance prediction can then be used with traditional kernel modules requiring accurate system state information. \aditya{rewrote the prev somewhat, but still the previous sentences don't connect well. Are we talking about predicting how the environment will evolve, or environment impact on app performance, or both.}

\noindent\textbf{Application behavior prediction:} Predicting the behavior of an application can help the OS prepare in advance for additional resources the application might need and minimize competition for shared hardware. For example, if the OS can predict that an application's execution will be memory-intensive in the near future based on its recent memory allocation calls and nature of inputs received, it can both provision more memory for the application and avoid scheduling another memory-intensive application on the same node.

\noindent\textbf{Anomaly Detection:} Using the state encoding of the OS or any of its components, and given a trace, one can ask if the current state is normal, or if there is some anomaly or failure issue.
Such predictions can be used to identify and kill anomalous applications, thereby improving the security of the OS kernel. %These require \abbr{} to be fine-tuned on diverse OS traces, with examples of anomalies.

%% file: texfiles/summary.tex
\section{Summary}

In conclusion, we argue that the OS decision-making tasks provide a rich arena for a domain-specific foundation model to be built for the OS.
We discuss the shortcomings of existing methods of data-driven decision-making and posit that rich OS traces can provide the necessary data to train such a foundation model, \abbr{}, which can understand the `natural behavior' of the OS.
We then provide a systematic analysis of the various ways in which \abbr{} can be used and the various key challenges that remain open research questions.

%% file: texfiles/appendix.tex
\appendix

\section{Decision Making Tasks in Operating Systems}\label{apdx:mdps}
\input{tables/os_components}

Table~\ref{tab:os_components} shows the various components in the OS and a \textit{representative subset} of the tasks for these components. It also shows the relevant system and environment states, action spaces, and the objectives of these tasks.
Each task description is also accompanied by an acronym that we use in the paper to refer to the particular task, e.g. \textbf{SCH} for the CPU scheduling task.

%% file: tables/os_components.tex
%%%%%%%%%%%%% TABLE VARIABLES %%%%%%%%%%%%%
\newcommand{\TableWidthColA}{0.9cm}
\newcommand{\TableWidthColB}{1cm}
\newcommand{\TableWidthColAB}{1.9cm}
\newcommand{\TableWidthColC}{3cm}
\newcommand{\TableWidthColD}{2.5cm}
\newcommand{\TableWidthColE}{2.25cm}
\newcommand{\TableWidthColF}{1.5cm}
\newcommand{\TableWidthColG}{1.5cm}
\newcommand{\TableVerticalSpacing}{1.2}
%%%%%%%%%%%%% TABLE VARIABLES %%%%%%%%%%%%%

\begin{table}[h]
\setlength\tabcolsep{2.5pt}
\tiny
% \begin{adjustwidth}{-1cm}{-1cm}
\centering
\renewcommand{\arraystretch}{\TableVerticalSpacing}
\caption{Various decision-making components in the OS.
% \ds{putting it early on in the paper to show what OS decisions we are talking about.}
% \todo{add GPU task}~\cjr{Regarding GPU tasks: we should discuss this: in principle a GPU is its own computer with multiple resources (compute, memory, PCIe bandwidth) which leads to instances of all the classic OS problems like scheduling, virtual memory management etc.. In practice, GPUs remain mostly not managed by single-host OSes such as Window or Linux, despite a wealth of research on how to do it. There is a ton of work on how to schedule jobs and low-level resources for GPUs at cluster level, but I think that might be out of scope for this paper. Probably we should take the position that GPUs will be OS-managed in future systems. }\aditya{we can drop GPUs without any impact on the paper, IMO.}
}
\label{tab:os_components}
\begin{tabular}
{|p{\TableWidthColA}p{\TableWidthColB}|p{\TableWidthColC}|p{\TableWidthColD}|p{\TableWidthColE}|p{\TableWidthColF}|p{\TableWidthColG}|}
\hline
\multicolumn{2}{|c|}{\textbf{Components}} & \multicolumn{1}{c|}{\textbf{Description and Acronym}} & \multicolumn{1}{c|}{\textbf{System State}} & \multicolumn{1}{c|}{\textbf{Environment Info}} & \multicolumn{1}{c|}{\textbf{Actions}} & \multicolumn{1}{c|}{\textbf{Objectives}} \\ \hline
\multicolumn{1}{|p{\TableWidthColA}|}{\multirow{2}{*}{\centering {CPU}}} & Scheduling & {\bf [SCH]:} Choose next process to run and which CPU to run it on~\cite{mllb-sigops-20} & Process state (niceness, priority, execution time), Hardware state (CPU, RAM spec, etc.) & Arrival pattern and type of processes (computation-heavy vs. I/O-heavy) & Process to core assignment & Job completion time, fairness \\ \cline{2-7} 
\multicolumn{1}{|c|}{} & Voltage and frequency scaling & {\bf [DVFS]:} Choosing CPU frequencies dynamically to reduce power consumptions~\cite{ztt-mobisys-21} & CPU frequency buckets, Hardware spec of the CPU & Process workloads, and process instructions & Choose CPU frequency & CPU performance, power, temperature \\ \hline
\multicolumn{1}{|p{\TableWidthColA}|}{\multirow{3}{=}{Memory Subsystem}} & Page Allocation & {\bf [ALLOC]:} What page size to use (e.g., huge pages vs normal pages) and how to allocate memory~\cite{llama-asplos-20} & Page table size, Hardware spec (amount of memory, type, etc.) & Memory access patterns of running processes & Page Size, Allocation mechanism & Latency of memory accesses \\ \cline{2-7} 
\multicolumn{1}{|c|}{} & Page Replacement & {\bf [PAGE]:} Choose a page in the physical memory to replace with another page in virtual memory~\cite{lpr-middleware-22} & Physical memory state, Hardware spec (amount of memory, type, etc.) & Program instructions, historical data of page faults for the processes & Choose the page to replace & Number of page faults \\ \hline
\multicolumn{1}{|p{\TableWidthColA}|}{\multirow{2}{=}{Network Subsystem}} & Packet Scheduling & {\bf [NETQUEUE]:} Order packets to send/process from the NIC queues & Queuing delays, NIC spec & Application type (video streaming, analytics, etc.) & Packet drop rate & Throughput and delay \\ \cline{2-7} 
% \multicolumn{1}{|c|}{} & Video Streaming & Generate adaptive bitrate (ABR) algorithms~\cite{pensieve-sigcomm-17} & Throughput samples, playback buffer occupancy, video chunk sizes & Bitrate decision for the next video chunk & Quality of Experience (QoE) & ms \\ \cline{2-7} 
\multicolumn{1}{|c|}{} & Congestion Control & {\bf [CC]:} Set congestion window, pacing rate for the connection~\cite{orca-sigcomm-20,rl-icml-19} & Network throughput, delay and packet loss; NIC spec & Application type (video streaming, analytics, etc.) & Congestion window, pacing rate & Throughput and delay \\ \hline
\multicolumn{1}{|p{\TableWidthColA}|}{\multirow{4}{=}{Storage Subsystem}} & I/O scheduling & {\bf [IOSCH]:} Deciding in which order I/O requests should be submitted to storage devices~\cite{linnOS-osdi-20} & I/O metadata (block offset, size), queue state, historical I/O latencies & Application type information (e.g., database, file system, etc.) & Order of I/O requests & I/O latencies \\ \cline{2-7} 
\multicolumn{1}{|c|}{} & Prefetching & {\bf [PREFETCH]:} Predict which segments of memory to prefetch~\cite{kml-hotstorage-21} & Cache size and state, Cache and PCIe spec & Process workloads and process instructions & Choose segment to prefetch & Throughput of future reads \\ \cline{2-7} 
\multicolumn{1}{|c|}{} & Cache replacement & {\bf [CACHE]:} Decide whether and which object to replace in the cache with the new object~\cite{darwin-sigcomm-23,lrb-nsdi-20,glcache-fast-23} & Cache size and state (occupied, address, last access) & Cache workloads (object sizes, frequency of access, etc.) & Choose a set of objects to evict/admit & Cache hit ratio\\ \hline
\end{tabular}
% \end{adjustwidth}
\end{table}